\documentclass{aa}               
\usepackage[dvips,final]{graphics}
\usepackage{natbib}
\usepackage{psfrag}

\psfrag{l}{$\lambda_{0}$}
\psfrag{W}{$\Omega_{0}$}

\begin{document}

\thesaurus{02(12.07.1; 12.03.4; 12.03.3; 12.03.2)}

\title{
Gravitational lensing statistics with extragalactic surveys
}

\subtitle{
III. Joint constraints on $\lambda_{0}$ and $\Omega_{0}$ from lensing 
statistics and the $m$-$z$ relation for type Ia supernovae
}

\author{
Phillip Helbig\inst{1,2}
} 

\institute{
University of Manchester, 
Nuffield Radio Astronomy Laboratories, 
Jodrell Bank, 
Macclesfield,
Cheshire SK11 9DL,
UK
\and 
University of Groningen,
Kapteyn Astronomical Institute, 
P.O. Box 800, 
9700 AV Groningen,
The Netherlands
}

\date{Received 23 April 1999 / Accepted 23 July 1999}

\offprints{P.~Helbig, Jodrell Bank address}
\mail{p.helbig@jb.man.ac.uk}

\authorrunning{P.~Helbig}
\titlerunning{
Gravitational lensing statistics with extragalactic surveys. III
}

\maketitle

\begin{abstract}
I present constraints on cosmological parameters in the
$\lambda_{0}$-$\Omega_{0}$ plane from a joint analysis of gravitational
lensing statistics \citep{PHelbigMQWBK99a} and the magnitude-redshift
relation for Type~Ia~supernovae
\citep{SPerlmutteretal99a,ARiessetal98a}.  I discuss reasons why this
particular combination of tests is important and how the constraints can
be improved in the future.  The lensing statistics and supernova results
are not inconsistent, thus it is meaningful to determine joint
constraints on $\lambda_{0}$ and $\Omega_{0}$ by combining the results
from both tests.  The quantity measured by the lens statistics and the
$m$-$z$ relation for type Ia supernovae discussed here is
approximately $\lambda_{0}-\Omega_{0}$.  At 95\% confidence, the upper
limit on $\lambda_{0}-\Omega_{0}$ from lensing statistics alone is
0.45 and from supernovae alone is in the range 0.65--0.81 (depending on
the data set).  For joint constraints, the upper limit on
$\lambda_{0}-\Omega_{0}$ is in the range 0.55--0.60 (again depending
on the data set).  For a flat universe with 
$\lambda_{0} + \Omega_{0} = 1$, this corresponds to upper limits on 
$\lambda_{0}$, taking the top of the range from different data sets, of 0.72,
0.90 and 0.80 for lensing statistics alone, supernovae alone and the 
joint analysis, respectively.  This is perfectly consistent with the 
current `standard cosmological model' with $\lambda_{0}\approx 0.7$ and 
$\Omega_{0}\approx 0.3$ and is consistent with a flat universe but, 
neglecting other cosmological tests, does 
not require it.

\end{abstract}

\section{Introduction}

Recently, several papers
\citep[e.g.][]{JOstrikerPSteinhardt95a,MTurner96a,JBaglaPN96a,LKrauss98a,
MWhite98a,MTegmarkEH98a,MTegmarkEHK98a,DEisensteinHT98a,
DEisensteinHT98b,MWebsterBHLLR98a,SBridleELLHCFH99a,GEfstathiouBLHE99a} have 
pointed out
the advantages of joint analyses of cosmological parameters,
i.e.~combining the information from more than one cosmological test. 
Ideally, such tests would be complementary, i.e.~the degeneracy in the
$\lambda_{0}$-$\Omega_{0}$ plane would be in orthogonal directions. 
However, even if this is not the case, indeed, even if the degeneracy is
exactly the same, the combination of tests can tighten the constraints
as well as serve as a consistency check.  Here, I discuss constraints on
cosmological parameters in the $\lambda_{0}$-$\Omega_{0}$ plane from a
joint analysis of gravitational lensing statistics
\citep[hereafter Paper~II]{PHelbigMQWBK99a} and the magnitude-redshift 
relation for
Type~Ia~supernovae, using the results of the Supernova Cosmology Project 
and the High-Z Supernova Search Team
\citep[hereafter SCP and HZSST, 
respectively]{SPerlmutteretal99a,ARiessetal98a}.  Although
both tests are preliminary in the sense that they will improve with more
and better observational data, the time is already ripe for a joint
analysis, to demonstrate both what already can be done and how each test
can be improved to lead to tighter joint constraints. 

The plan of this paper is as follows.  In Sect.~\ref{theory} I briefly
review the basis of each of these two cosmological tests.  In
Sect.~\ref{results} I present and discuss the joint constraints.
Sect.~\ref{conclusions} provides a summary and conclusions.

\section{Theory review}
\label{theory}

I here use the notation of \citet{RKayserHS97a} with regard to cosmology
and refer the reader there for the relevant definitions.  In particular,
$\Omega_{0}$ refers only to `ordinary matter' and $\lambda_{0}$ is the
normalised cosmological constant, such that $\lambda_{0}+\Omega_{0}=1$
for a flat universe. 

Both gravitational lensing statistics and the magnitude-redshift
relation are `classical' cosmological tests, i.e.~the theoretical
dependence of an observable quantity on redshift is compared with
observations.  This is done straightforwardly in the case of the
magnitude-redshift relation, and in a somewhat more roundabout way in
the maximum-likelihood analysis of gravitational lens statistics used
here.  The redshift range probed by the magnitude-redshift relation
extends at present out to $z \approx 1$.  In the case of lensing
statistics, the source population extends to quite large redshifts ($z
\approx 4$) although the redshift range of significant optical depth is
smaller.  Thus, the two tests are both `global' rather than `local'
cosmological tests and probe similar, though not identical, redshift
ranges.  Otherwise, the tests are completely independent. 

The $m$-$z$ relation is concerned essentially only with the luminosity
distance $D^{\mathrm{L}}$ whereas lensing statistics deal with several 
different angular
size distances (between observer and lens ($D_{\mathrm{d}}$), observer
and source ($D_{\mathrm{s}}$) and lens and source ($D_{\mathrm{ds}}$))
\citep[see, e.g.,][for a discussion of the various cosmological 
distances]{RKayserHS97a} and the volume; they also depend on several other
`astrophysical' parameters
\citep[e.g.][hereafter Paper~I]{CKochanek96a,RQuastPHelbig99a}.

\subsection{The $m$-$z$ relation for type Ia supernovae}

The basic idea of the $m$-$z$ relation is simple: one has an object of 
known absolute magnitude $M$ and compares it to the observed 
magnitude $m$.  The difference or distance modulus is
\begin{equation}
\label{eq:m}
m - M = 5\log_{10} D^{\mathrm{L}} + K + 42.384 - 5\log_{10}h ,
\end{equation}
where $D^{\mathrm{L}}$ is in units of the Hubble length, $K$ is the 
$K$-correction and $h$ is the Hubble constant in units 
of $100$ \mbox{km/s/Mpc} 
\citep[see, e.g.,][for a derivation]{RKayserHS97a}\footnote{Note that
the second occurrence of the term `Hubble length' in \citet{RKayserHS97a} 
should actually be `Hubble length for $h=1$', although this is obvious 
from the context.}.  This depends on the cosmological model since 
$D^{\mathrm{L}}$ depends on the cosmological parameters $\lambda_{0}$ 
and $\Omega_{0}$.  Note that, as is often the case in practice, if $M$ 
is known modulo $h$, then Eq.~(\ref{eq:m}) does not depend on the Hubble 
constant at all.  On the other hand, if $M$ is known absolutely, this is 
equivalent to knowing $h$, assuming one has at least one object at low 
redshift (where the dependence on $\lambda_{0}$ and $\Omega_{0}$ is 
negligible).  In any case, our knowledge (or lack of it) about the value 
of the Hubble constant $H_{0}$ does not appreciably affect the ability 
of this cosmological test to measure the cosmological constant
$\lambda_{0}$ and the density parameter $\Omega_{0}$.

Thus, one has a number of objects with observed magnitudes $m_{i}$ and a 
way of calculating the absolute magnitudes $M_{i}$ 
(see, e.g., SCP and HZSST for a description of how this is done in
practice)
and fits for the parameters $\lambda_{0}$ and $\Omega_{0}$.  If all 
objects are in a narrow redshift range, then confidence contours in the 
$\lambda_{0}$-$\Omega_{0}$ plane will only allow one to measure
approximately $\lambda_{0}-\Omega_{0}$ whereas having objects at 
different redshifts breaks this degeneracy 
\citep[e.g.][]{AGoobarSPerlmutter95a}.

\subsection{Gravitational lensing statistics}

See Paper~I and references therein for a discussion 
of how constraints on $\lambda_{0}$-$\Omega_{0}$ are derived from 
gravitational lensing statistics.   Gravitational lensing statistics, at 
least in the `interesting' part of parameter space, constrain approximately 
$\lambda_{0} - \Omega_{0}$ \citep[e.g.][]{ACooray99a}.  Thus the degeneracy is 
approximately the same as that of the $m$-$z$ test.  Thus, rather than 
reducing the allowed area of parameter space through orthogonal 
degeneracies, these two cosmological tests provide a 
consistency check on each other.  Also, the $m$-$z$ relation provides a 
good \emph{lower} limit on $\lambda_{0}$ while lensings statistics 
provides an \emph{upper} limit; obviously, the former should be smaller 
than the latter.  If this is the case, then the two cosmological tests 
are consistent with each other, and it is meaningful to construct joint 
constraints, which allow a region of parameter space smaller than that 
allowed by either test alone.

\section{Data and results}
\label{results}

\subsection{Individual results}

For the $m$-$z$ test I used the results presented in
SCP and HZSST, which have kindly
been made available by the respective collaborations, as well as our own
results from the analysis of JVAS, the Jodrell Bank-VLA Astrometric Survey  
(Paper~II and references therein).
Fig.~\ref{fi:supernovae} shows the likelihood ratio as a grey scale and
the 68\%, 90\%, 95\% and 99\% confidence contours for the results from
SCP and HZSST; Fig.~\ref{fi:lenses} 
does the same for the results from Paper~II.
\begin{figure*}
\noindent
\resizebox{0.3\textwidth}{!}{\includegraphics{8785.f1}}
\hfill
\resizebox{0.3\textwidth}{!}{\includegraphics{8785.f2}}
\hfill
\resizebox{0.3\textwidth}{!}{\includegraphics{8785.f3}}
\vspace{1ex}
\noindent
\resizebox{0.3\textwidth}{!}{\includegraphics{8785.f4}}
\hfill
\resizebox{0.3\textwidth}{!}{\includegraphics{8785.f5}}
\hfill
\resizebox{0.3\textwidth}{!}{\includegraphics{8785.f6}}
\vspace{1ex}
\noindent
\resizebox{0.3\textwidth}{!}{\includegraphics{8785.f7}}
\hfill
\resizebox{0.3\textwidth}{!}{\includegraphics{8785.f8}}
\hfill
\resizebox{0.3\textwidth}{!}{\includegraphics{8785.f9}}
\caption[]{The likelihood function $p(D|\lambda_{0},\Omega_{0})$ (cf.~Paper~I) 
from
\citet{SPerlmutteretal99a} (SCP, equivalent to their Fig.~7; hereafter data 
set $\cal A$) (left column), 
\citet{ARiessetal98a} (HZSST) ($\Delta m_{15}(B)$ method, equivalent
to the dotted contours of 
their Fig.~7; hereafter data set $\cal B$) (middle column) and 
\citet{ARiessetal98a} (MLCS method, equivalent to the 
dotted contours of their Fig.~6; hereafter data set $\cal C$) (right
column) in the original parameter space and
resolution (top row), in the parameter space used for the calculations
in this paper but in the original resolution (middle row) and in the
parameter space used for calculations in this paper in the resolution
used for calculations in this paper (bottom row).  The pixel grey level
is directly proportional to the likelihood ratio, darker pixels reflect
higher ratios. The contours mark the boundaries of the minimum $0.68$,
$0.90$, $0.95$ and $0.99$ confidence regions for the parameters
$\lambda_{0}$ and $\Omega_{0}$} 
\label{fi:supernovae}
\end{figure*}
\begin{figure*}
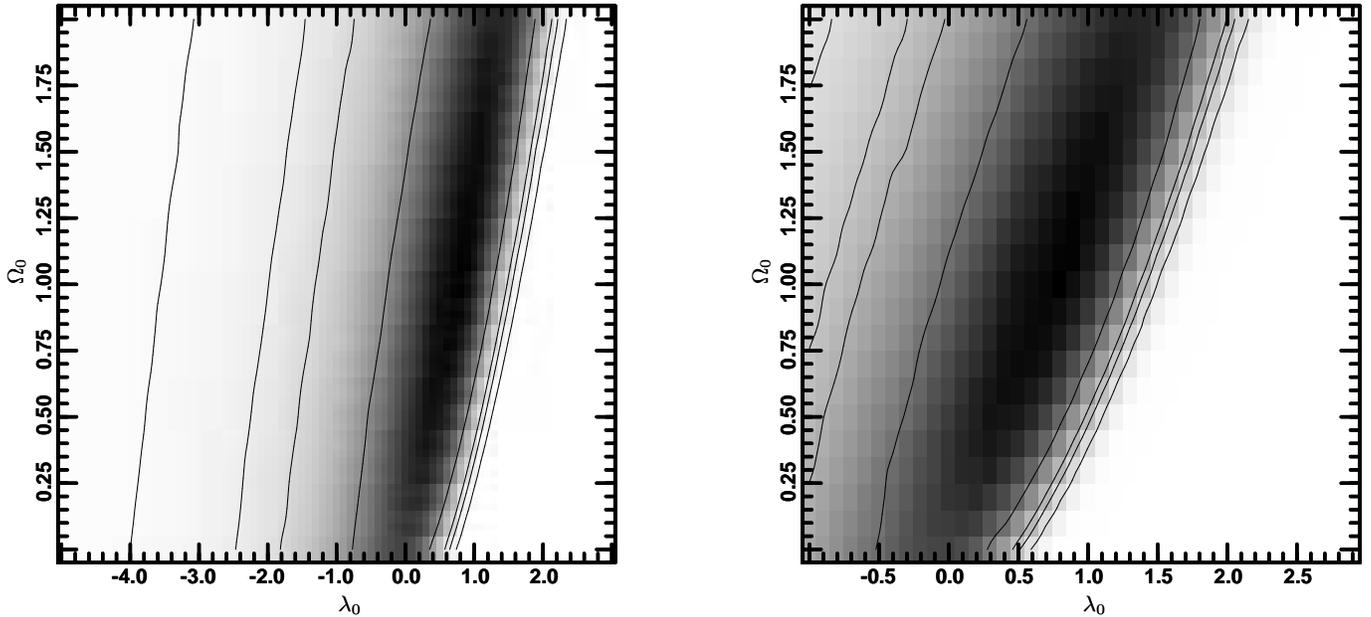

\noindent
\resizebox{0.45\textwidth}{!}{\includegraphics{8785.f10}}
\hfill
\resizebox{0.45\textwidth}{!}{\includegraphics{8785.f11}}
\caption[]{The likelihood function $p(D|\lambda_0,\Omega_0)$ from
Paper~II (hereafter data set $\cal D)$ in the original parameter space 
(left) and in the parameter space used for the calculations
in this paper (right).  (The resolution in Paper~II is 
(due to the fact that the lens statistics calculations are numerically 
much more demanding) the worst of all the data sets considered here 
and is thus the one used for the calculations in this paper.)  See 
Fig.~\ref{fi:supernovae} for a description of the plotting scheme} 
\label{fi:lenses}
\end{figure*}
See these references for discussions of these results individually.  I
use the JVAS results of Paper~II, rather than those of
Paper~I,
since the former seem more reliable, despite the remaining uncertainties
(see Paper~II for a discussion).  Also, using only one
set of lensing statistics results, rather than a combination, is
conservative, since the joint constraints are tighter than individual
constraints.\footnote{Of course, if one is concerned with the
consistency of the results, rather than in reducing the parameter space
through joint constraints, then one should use as many results
as possible.  However, it only makes sense in this context to use
reliable results, so this is a reason to neglect the results based on
optical gravitational lens surveys discussed in
Paper~I.} 

The basic format here is that of a probability density function, i.e.~a
relative probability as a function of $\lambda_{0}$ and $\Omega_{0}$.
Ideally, this would cover \emph{all} values of $\lambda_{0}$ and
$\Omega_{0}$, or at least all for which there is a non-negligible
probability.  Alternatively, one can impose a prior constraint on $\lambda_{0}$,
$\Omega_{0}$ or both, such that there is a non-negligible probability
only in a comfortably small region of parameter space.  The simplest way
to do this is to use a top-hat function, such that the a posteriori
likelihood is given by the a priori likelihood within some range and
is exactly zero outside of this range.  This is a conservative approach
if the allowed range is large enough to include the correct
cosmological model in any case and also since, within the allowed range,
the likelihood depends only on the cosmological tests considered and not
on the priors (which makes for easier interpretation). 

The first three rows of Table~\ref{ta:ranges} show the range of parameter
space covered by the references which are used to provide input data for
this work. 
\begin{table*}
\caption[]{The range of $\lambda_{0}$ and $\Omega_{0}$ explored by the 
references used here}
\label{ta:ranges}
\begin{tabular*}{\textwidth}{@{\extracolsep{\fill}}lrrcrrc}
\hline
Reference
& 
\multicolumn{3}{c}{$\lambda_{0}$}
& 
\multicolumn{3}{c}{$\Omega_{0}$}
\\
&
\multicolumn{2}{c}{range}
&
resolution
&
\multicolumn{2}{c}{range}
&
resolution
\\
\hline
\citet{SPerlmutteretal99a} (SCP)
&
-1.00
&
2.98
&
0.02
&
0.00
&
2.99
&
0.01
\\
\citet{ARiessetal98a} (HZSST)
&
-1.00
&
3.00
&
0.01
&
0.00
&
4.00
&
0.01
\\
\citet{PHelbigMQWBK99a} (Paper~II)
&
-5.00
&
3.00
&
0.10
&
0.00
&
2.00
&
0.10
\\
this work
&
-1.00
&
2.90
&
0.10
&
0.00
&
2.00
&
0.10
\\
\hline
\end{tabular*}
\end{table*}
I take as the only prior that the likelihood is zero outside the overlap
of the various ranges of the various cosmological tests used, as in the
last row in Table~\ref{ta:ranges}.  The lower limit on $\Omega_{0}=0$ is
physical and the upper limits $\Omega_{0}=2$ and $\lambda_{0}=2.9$ are
certainly large enough (see the discussion in Paper~I on these values and
on the use of prior information in general). The lower
limit on $\lambda_{0}$ comes mainly from the fact that $\lambda_{0}<-1$
is strongly excluded by the $m$-$z$ relation itself, although the
analysis of current cosmic microwave background observations
\citep[e.g.][]{CLineweaver98a,JPerezHQ99suba,PHelbigBBdBFJKMMMQRWX98_Ca}
also suggests this.  In any case, an analysis of joint constraints from
other cosmological tests, even excluding the $m$-$z$ relation, suggests
that $\lambda_{0}>0$ is a robust result \citep[e.g.][]{MRoosSHoR99a}. 

With an individual test, likelihood contours are found (in all of the
cases discussed here; see the discussion in Paper~I for
possible caveats when comparing the results of various cosmological
tests as presented in the literature) by finding the highest contour at 
constant likelihood 
such that the corresponding fraction of the total likelihood is
enclosed.  Note that these contours depend not only on the likelihood
ratio but also on the range of parameter space plotted.  In this work,
the parameter space plotted should be considered to have an a priori
likelihood of 1, while the parameter space outside the plot should be
considered to have an a priori likelihood of 0.  Note that these are
`real' likelihood contours, not approximations based on
$\Delta\chi^{2}$, the assumption that the probability distribution is 
(a 2-dimensional) Gaussian etc such as one often finds in the 
literature.

It should also be noted that I take the likelihood as presented in the
references in question.  In the case of the lens statistics (data set 
$\cal D$), all
parameters except $\lambda_{0}$ and $\Omega_{0}$ were held constant.  In
the case of the $m$-$z$ relation (data sets $\cal A$--$\cal C$), the 
results are
obtained by marginalising over the nuisance parameters (see,
e.g., Paper~I for a discussion).  However, this is of no
concern at the level of accuracy I am concerned with here, especially
since there are no nuisance parameters common to the $m$-$z$ test and
the lensing statistics test.\footnote{The publicly available data from
SCP are actually not the likelihood itself, but
rather the value for each point in the parameter space is the
(normalised) sum of all likelihood values for all points in the 
parameter space which are not less than the value for the point in 
question.  This format, which allows one to immediately plot a given 
confidence contour by plotting a contour at that level, I have converted 
back to the original probability density function.}

\subsection{Joint constraints}

\begin{figure*}
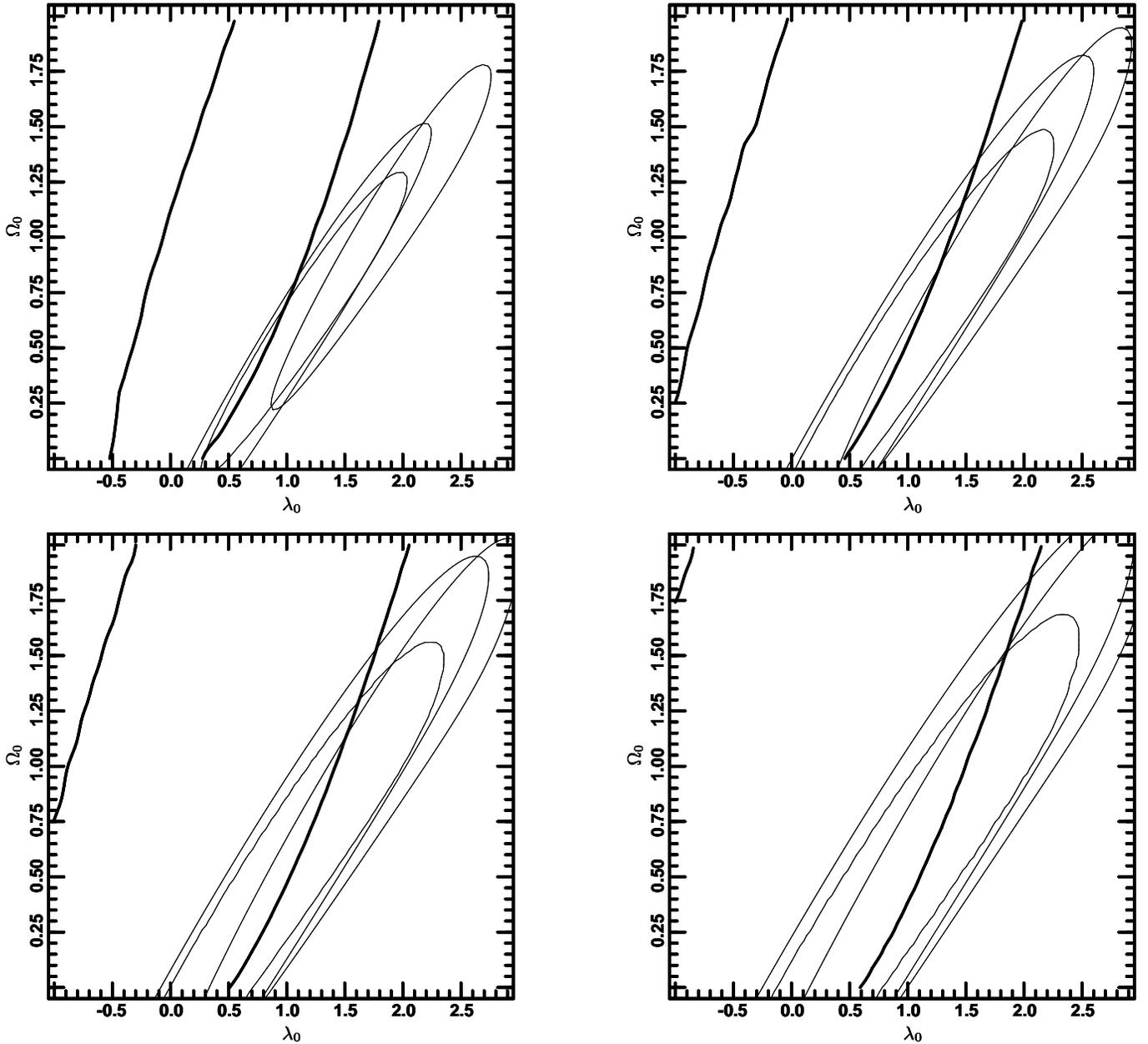

\noindent
\resizebox{0.45\textwidth}{!}{\includegraphics{8785.f12}}
\hfill
\resizebox{0.45\textwidth}{!}{\includegraphics{8785.f13}}

\vspace{1.5ex}

\noindent
\resizebox{0.45\textwidth}{!}{\includegraphics{8785.f14}}
\hfill
\resizebox{0.45\textwidth}{!}{\includegraphics{8785.f15}}
\caption[]{The 68\% (top left), 90\% (top right), 95\% (bottom left) and 
99\% (bottom right) confidence contours for each of the data sets.  The 
thick curves are for the lensing statistics results (data set $\cal D$).
In all plots, data set $\cal A$ has the contour with the lowest value of 
$\Omega_{0}$ at its maximum height.  Starting from this point and moving 
left, towards smaller values of $\lambda_{0}$, in all plots one crosses 
first the contour of data set $\cal B$ then that of data set $\cal C$}
\label{fi:overlap}
\end{figure*}

\begin{figure*}
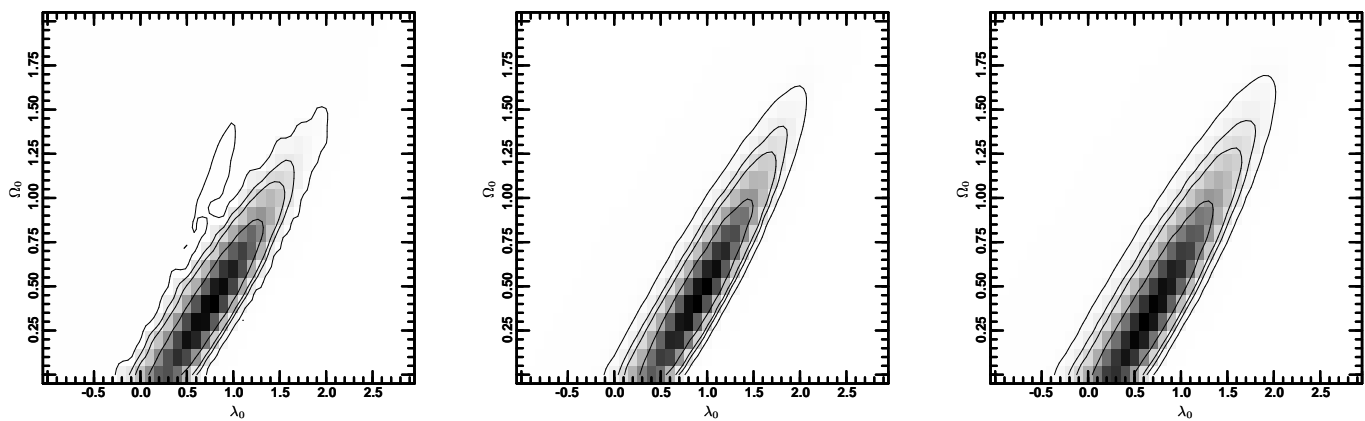

\noindent
\resizebox{0.3\textwidth}{!}{\includegraphics{8785.f16}}
\hfill
\resizebox{0.3\textwidth}{!}{\includegraphics{8785.f17}}
\hfill
\resizebox{0.3\textwidth}{!}{\includegraphics{8785.f18}}
\caption[]{Joint constraints from the lensing statistics calculations of
\citet{PHelbigMQWBK99a} (Paper~II) 
and the $m$-$z$ relation for type Ia supernovae as in 
data set $\cal A$ (left), 
data set $\cal B$ (middle) and 
data set $\cal C$ (right).  See 
Fig.~\ref{fi:supernovae} for an explanation of the data sets}
\label{fi:joint}
\end{figure*}

The simplest thing to do when building joint constraints would be to
multiply the corresponding probability density functions
(PDFs).\footnote{Of course, this must be done at the same resolution.
Rather than interpolate the low resolution lens statistics results, I
have reduced the resolution of the $m$-$z$ results to that of the
lensing statistics results by using only those points in the
$\lambda_{0}$-$\Omega_{0}$ plane which were examined in the lens
statistics calculations, all of which were examined by both $m$-$z$
tests.}  One can then plot confidence contours in the manner described
above.  However, it is obvious that this is not meaningful if the PDFs
are not consistent with each other, i.e.~if the region of confidence for
a `sensible' confidence level from the joint constraints does not
overlap with the corresponding confidence level for all component tests.
A necessary, though not sufficient, condition for this inconsistency to
exist is that the corresponding confidence contours for the individual
component tests do not overlap.  Fig.~\ref{fi:overlap} shows the 68\%,
90\%, 95\% and 99\% confidence contours for the four data sets 
considered here.  
As the 90\% confidence contours from all supernovae data sets overlap
with that of the lensing statistics, and even the 68\% confidence
contours from two of three supernovae data sets overlap with that of the
lensing statistics, the results from the two cosmological tests are 
consistent and one is justified in calculating joint constraints by 
multiplying the probability distributions of the individual 
tests.  Interestingly, they are most consistent at small, but not too 
small, values of $\Omega_{0}.$  The results of this are shown in 
Fig.~\ref{fi:joint}.

Note that if there is some offset between the allowed regions from each 
individual test, as is the case here, then certain aspects of the joint 
constraints, such as in this case the upper limit on $\lambda_{0}$, will 
not necessarily be tighter than the corresponding aspect from each 
individual test.  The joint constraints are nevertheless better in that 
the allowed region is smaller and that this allowed region should 
contain the correct cosmological model, assuming of course that the 
results of the individual tests are correct as far as they go.  The 
tests are very different in nature and one should not expect the form of 
the probability density function to be the same in each case.  In 
particular, as lensing statistics is especially sensitive to a large 
cosmological constant, the gradient in this area of parameter space is 
quite steep, thus it is not surprising that the lensing statistics upper 
limit on $\lambda_{0}$ is tighter.  The fact that the confidence 
contours from the individual tests overlap shows that the tests are not 
inconsistent, and of course the allowed region from the joint 
constraints, which is consistent with each individual test, is 
approximated by this overlap.

Since the two $m$-$z$ results are not completely independent, the
question of the consistency of or joint constraints from the two
supernovae data sets will not be discussed in this paper.  Rather, the
question is the consistency of and joint constraints from each of these
data sets individually with the lensing statistics constraints. 

Fig.~\ref{fi:joint} is the main conclusion of this paper.  Although 
lensing statistics and
the $m$-$z$ relation individually allow, for appropriate values of
$\lambda_{0}$, rather large values of $\Omega_{0}$, the joint
constraints clearly indicate a lower $\Omega_{0}$, in accordance with
observational evidence which measures $\Omega_{0}$ more `directly' (see
the discussion in Paper~I).  Compared to the supernovae results, the
allowed region of parameter space is shifted somewhat towards lower
values of $\Omega_{0}$ in the joint constraints.  Although the actual
best-fit value should not be taken too seriously, it is comfortably
close to the current `standard cosmological model' with
$\lambda_{0}\approx 0.7$ and $\Omega_{0}\approx 0.3$ 

The quantity measured by both the lens statistics and the
$m$-$z$ relation for type Ia supernovae discussed here is
approximately $\lambda_{0}-\Omega_{0}$.  
Table~\ref{ta:results} shows the 95\% confidence ranges for
$\lambda_{0} - \Omega_{0}$ allowed by each 
of the four data sets individually and by the joint constraints of data 
set $\cal D$ with data sets $\cal A$, $\cal B$ and $\cal C$.
\begin{table*}
\caption[]{95\% confidence ranges for
$\lambda_{0} - \Omega_{0}$ allowed by each 
of the four data sets individually as well as various joint constraints}
\label{ta:results}
\begin{tabular*}{\textwidth}{@{\extracolsep{\fill}}lcc}
\hline
Reference
& 
lower limit
& 
upper limit
\\
\hline
\citet{SPerlmutteretal99a} (SCP) (data set $\cal A$)
&
$-0.05$
&
$+0.65$
\\
\citet{ARiessetal98a} (HZSST, $\Delta m_{15}(B)$) (data set $\cal B$)
&
$+0.30$
&
$+0.81$
\\
\citet{ARiessetal98a} (HZSST, MLCS) (data set $\cal C$)
&
$-0.12$
&
$+0.79$
\\
\citet{PHelbigMQWBK99a} (Paper~II) (data set $\cal D$)
&
$-1.90$
&
$+0.45$
\\
$\cal A$ + $\cal D$
&
$-0.20$
&
$+0.55$
\\
$\cal B$ + $\cal D$
&
$\pm 0.00$
&
$+0.55$
\\
$\cal C$ + $\cal D$
&
$-0.25$
&
$0.60$
\\
\hline
\end{tabular*}
\end{table*}
At 95\% confidence, the upper
limit on $\lambda_{0}-\Omega_{0}$ from lensing statistics alone is
0.45 and from supernovae alone is in the range 0.65--0.81 (depending on
the data set).\footnote{Note that this is not the same as that quoted in 
Paper~II; this is because, as discussed above, the range of parameter 
space examined or, equivalently (in this case), the prior is different.}  
For joint constraints, the upper limit on
$\lambda_{0}-\Omega_{0}$ is in the range 0.55--0.60 (again depending
on the data set).  For a flat universe with 
$\lambda_{0} + \Omega_{0} = 1$, this corresponds to upper limits on 
$\lambda_{0}$, taking the top of the range from different data sets, of 0.72,
0.90 and 0.80 for lensing statistics alone, supernovae alone and the 
joint analysis, respectively.  Again, this is perfectly consistent with the 
current `standard cosmological model' with $\lambda_{0}\approx 0.7$ and 
$\Omega_{0}\approx 0.3$ \citep[e.g.][]{MRoosSHoR99a,MTurner99a}
and is consistent with a flat universe but, 
neglecting other cosmological tests, does 
not require it.

\subsection{Systematic errors}

As far as the $m$-$z$ relation for type Ia supernovae goes, various
possible sources of systematic errors have been discussed in detail by
SCP and HZSST.  See particularly Fig.~5 in SCP.  Basically, there is
no evidence that the purported effects could significantly bias the
results or, as in the case of grey dust, while a modest amount cannot be
ruled out, it seems physically rather implausible \citep[but see,
however,][]{AAguirre99a, AAguirre99b,AAguirreZHaiman99a}.  (Note that
\citet{EFalcoIKLMcLRKMP99a} find no evidence for grey dust at optical
wavelengths based on studies of extinction in gravitational lens
galaxies.)  It is interesting to note that while one can invoke grey dust
to explain the dimming of supernovae relative to the expectation in for
example a dust-free $\lambda_{0}=0$ model, instead of invoking for
example a low-density model with a positive cosmological constant, this
degeneracy can be broken by observing supernovae at higher redshift
\citep[e.g.][]{AAguirre99a} than has been done up until now: the $m$-$z$
relation as a function of $\lambda_{0}$ and $\Omega_{0}$ is exactly
known, so the larger the range in redshift for which the $m$-$z$
relation is observed, the more ad hoc alternative explanations become,
provided of course that there is a cosmological model (which is not
ruled out on other grounds) which provides an acceptable fit to the
data. 

Grey dust is also something which can effect gravitational lensing
statistics \emph{based on optical samples}, although the effects are not
so straightforward.  On the one hand, if the grey dust is concentrated
in (lensing) galaxies, this could lead to lens systems being missed in
the survey.  To first order, this would lead to an underestimate of the
optical depth and thus of the value of $\lambda_{0}$.  On the other
hand, again if dust is concentrated in (lensing) galaxies, the sources
in the identified lens systems can suffer from extinction, which,
depending on the details of the luminosity function, could lead to a
wrong estimate of the magnification bias.  In the `normal' case of a
flattening of the luminosity function for fainter objects, this will
lead to an underestimate of the amplification bias and hence an
overestimate of the optical depth and thus the value of $\lambda_{0}$
\citep{EFalcoIKLMcLRKMP99a}.  Radio surveys of course are not affected by
dust, so in principle one could detect grey dust through a systematic
difference in the results from optical and radio surveys.  In practice,
however, the presence of other systematic effects makes such a detailed
comparison impractical. 

Radio surveys for gravitational lenses offer many advantages over
optical surveys (see the discussion in Paper~II).  However, at present,
the main source of uncertainty, lack of knowledge about the source
population, makes them worse than optical surveys in this respect.  For
the calculation of the amplification bias, one needs to know, at a given
redshift, the luminosity function.\footnote{Due to the amplification of
the gravitational lens effect, lensed sources near the lower
flux-density limit of the survey will have an unlensed flux density
lower than this, so the luminosity function thus needs to be known down
to a flux-density level a factor of several below that of the survey.}
On the other hand, much information is gained from the sources in a
survey which are not lensed (see the discussion in Paper~I); to
interpret this, one needs to know, at a given flux density, the redshift
distribution of the sources.  Of course, these two things---the
redshift-dependent luminosity function and the flux-density dependent
redshift distribution---are different sides of the same coin. 

The number counts of the Cosmic Lens All-Sky Survey, of which the
Jodrell Bank-VLA Astrometric Survey, the results of which are used here,
is a subset, suggest that amplification bias is not a big effect; hence
the systematic error from lack of knowledge of the luminosity function
is probably small, although it is conceivable that the number counts
(integrated over redshift, as in general the CLASS sources have unknown
redshifts) of CLASS are not representative of the luminosity function at
\emph{all} redshifts.  As the lensed sources are generally at higher
redshifts, the luminosity function might be different here and thus the
true amplification bias different from that which was used in the JVAS
analysis of Paper~II (where it was assumed that the CLASS number counts
are representative of all redshifts). 

In Paper~II, it was also assumed that the redshift distribution of JVAS
is equal to that of CJF, independent of flux-density.  There is some
preliminary evidence that, as one moves toward lower flux-density
levels, the typical redshift of flat-spectrum radio sources decreases.
If this is the case, then our JVAS analysis will have underestimated the
value of $\lambda_{0}$, as a higher value of $\lambda_{0}$ (all other
things being equal) is needed to achieve the same optical depth for a
low-redshift source than is needed for a high-redshift source.  Although
the results from the $m$-$z$ relation for type Ia supernovae and
gravitational lensing statistics are not inconsistent, and although, due
to the different dependence on the cosmological parameters, the lower
limit on $\lambda_{0}$ will always be stronger from the former and the
upper limit from the latter, if it does turn out to be true that the CJF
redshift distribution is systematically higher than the true redshift
distribution of JVAS, then the results from the $m$-$z$ relation for
type Ia supernovae and gravitational lensing statistics will become even
more consistent.

\section{Summary and conclusions}
\label{conclusions}

I have presented the first detailed analysis of joint constraints
between gravitational lensing statistics and the $m$-$z$ relation for
type Ia supernovae, making use of data from \citet{PHelbigMQWBK99a},
\citet{SPerlmutteretal99a} and \citet{ARiessetal98a}, presenting the
individual results and the new joint constraints in a uniform way.  The
two tests are not inconsistent, the joint constraints are tighter than
those from either test individually and provide additional evidence in
favour of the current `standard cosmological model' with
$\lambda_{0}\approx 0.7$ and $\Omega_{0}\approx 0.3$, although 
(neglecting constraints from other sources such as the CMB) a
reasonable range of other cosmological models is not excluded.

In the near future, gravitational lensing statistics from CLASS, the Cosmic 
Lens All-Sky Survey 
\citep{SMyersetal99a} should reduce both the random and systematic 
errors.  Should the results from lensing statistics and the 
$m$-$z$ relation for type Ia supernovae remain consistent, this should 
reduce the allowed parameter space even further.  We are truly entering 
an era of precision cosmology, where the overlap of the allowed regions 
of parameter space from many different and independent cosmological 
tests is very small but not zero.

The data for the figures shown in this paper are available at
\begin{quote}
\verb|http://multivac.jb.man.ac.uk:8000/ceres|\\
                                       \verb|/data_from_papers/snlens/snlens.html|
\end{quote}
or
\begin{quote}
\verb|http://gladia.astro.rug.nl:8000/ceres|\\
                                       \verb|/data_from_papers/snlens/snlens.html|
\end{quote}

\begin{acknowledgements}
It is a pleasure to thank Saul Perlmutter, Brian Schmidt and Saurabh Jha
for helpful discussions and the Supernova Cosmology Project and the
High-Z Supernova Search Team for making their numerical results
available.  This research was supported in part by the European
Commission, TMR Programme, Research Network Contract ERBFMRXCT96-0034
`CERES'. 
\end{acknowledgements}

\bibliographystyle{aa}

\end{document}